# A Creativity Survey of Parallel Sorting Algorithm

Tianyi Yu

School of Computer Science and Technology, Xidian University

yutianyi@stu.xidian.edu.cn

Wei Li

Institute of Computing Technology, CAS

liwei@ict.ac.cn

***Abstract:*** *Sorting is one of the most fundamental problems in the field of computer science. With the rapid development of manycore processors, it shows great importance to design efficient parallel sort algorithm on manycore architecture. This paper studies the parallel memory sorting method on modern hardware, and summarizes its research status and progress. Classify the research problems, research methods and measurement methods of the target papers and references. In the end, we summarize all the researches and list the directions not researched and innovative places.*

## 1 INTRODUCTION

With the rapid development of computer hardware technology, the computing power of computer is continuously improved. On the one hand, the processor structure is constantly changing. From the traditional single-core processor to multi-core processor, and then to manycore processor, the number of processor cores is increasing. On the other hand, with the continuous development of storage technology and the continuous increase of storage methods, from external memory to memory and then to cache, the memory access efficiency of computer is continuously improved. The development of computer hardware technology continues to shorten the response time of programs. In addition to the traditional CPU architecture, the emergence of new processor devices makes high-performance computing enter a new era.

Sorting is an essential basic data processing operation in many computer fields. There are a large number of sorting tasks in data mining, super computing, information systems and other fields. Sorting can make the data arranged in a certain order, so as to reduce the time of subsequent operations, For example, data search operations are carried out on ordered data sets. Sorting not only enables data to be accessed sequentially in memory, and the memory access efficiency is higher than random access, but also uses efficient data search methods such as binary search and search by displacement, so that the search efficiency of ordered data sets is much higher than that of unordered data sets, Therefore, the performance optimization of sorting has been concerned by researchers. In the field of sorting, many researchers have proposed many sorting methods. These methods have their own advantages and disadvantages, and also have their own suitable application scenarios. The development of computer hardware conditions also puts forward new requirements for the optimization of sorting methods. According to the latest computer hardware conditions and the application of sorting algorithm, researchers have done corresponding optimization work on CPU, GPU of many-core architecture and field programmable gate array (FPGA) focusing on hardware design, which has greatly improved the performance of sorting operation.

In the early days, due to the shortage of computer memory, for quite a long time, researchers' study on sorting operation mainly focused on external memory sorting. The data is stored in external memory devices. The program needs to read the data from external memory devices to internal memory for sorting. In the process of sorting, there will be many exchanges of intermediate results between external memory and memory, Therefore, the main performance bottleneck of external memory sorting lies in the frequent I / O access between memory and external memory. With the increase of memory capacity, all data can be stored in memory and sorted, and the I / O bottleneck between internal and external memory disappears, In addition, the increase of number of single processor's core, the concept of multithreading and the maturity of software parallel technology have continuously improved the parallel processing ability of processor devices, The performance of the sorting method implemented on it has also been greatly improved. The same as the problems faced by other parallel computing, the parallel sorting algorithm needs to avoid the problems of load imbalance, multi-threaded resource contention, memory access conflict and so on, The emergence of new computing processor devices also provides a new optimization direction for sorting methods

The rest of the paper is organized as follows. Section II gives the classification of research objects of parallel sorting algorithm. Section III introduces the classification of research methods. Section IV introduces the comparison of experimental analysis in related literature. Section V discusses the research opportunities in future work and Section VI concludes the paper.

## 2  CLASSIFICATION OF RESEARCH OBJECTS

Table 1: Different Research Objects

| Parallel Sorting Algorithm | Platform | | |
|---|---|---|---|
| | CPU | GPU | Others |
| Merge Sort | I. [1][3][11][14][21] | II. [2][5][7][12][13] | III. [0][10] |
| Radix Sort | IV. [19] | V. [15] | VI. |
| Others | VII. [16][17][20] | VIII. [4][6][9][18] | IX. [8] |

## 2.1 Criteria

Due to the great difference of different sorting methods, they can be divided into different research objects. In addition, there are many implementation platforms for the algorithm, which are mainly divided into three types: CPU, GPU and Others, among which Others include FPGA, hybrid platform, or platform independent and so on. In this section, two independent and different criteria would be used to divide research objects into different types:

1) **Sorting Algorithm**. There are three types here: **Merge Sort**, **Radix Sort** or **Others**. Different kinds of sorting algorithms are suitable for different kinds of situations. Sometimes they need to save space and have less time requirements. On the contrary, sometimes they want to consider more time and have less space requirements. In short, they generally have to make a choice from one aspect.

2) **Platform**. There are three kinds of algorithm implementation platform here: **CPU**, **GPU** or **Others**. There may be a variety of hardware available to people. By selecting algorithms on different platforms to meet the needs of sorting operation on different hardware.

## 2.2 The Classification

Based on the appeal classification standard, we give the classification in Table 1. The meaning of each class is as follows:

**Type I.** This type is parallel merge sort algorithm on CPU platform.

**Type II.** This type is a parallel merge sort algorithm on the GPU platform.

**Type III.** This type is a parallel merge sort algorithm on other platforms.

**Type IV.** This type is the parallel radix sorting algorithm on the CPU platform.

**Type V.** This type is the parallel radix sorting algorithm on the GPU platform.

**Type VI.** This type is the parallel radix sorting algorithm on other platforms.

**Type VII.** This type is other parallel sorting algorithms on the CPU platform.

**Type VIII.** This type is other parallel sorting algorithms on the GPU platform.

**Type IX.** This type is other parallel sorting algorithms on other platforms.

Explanation of Different Types

References ([1][3][11][14][21]) belong to Type I. The classical version of parallel merge sorting algorithm was first proposed in reference [3]. Reference [1][11][14][21], These all focus on the parallel merging and sorting algorithm on CPU.

References ([2][5][7][12][13]) belong to Type II. Reference [2] is associated with reference [1] focusing on CPU, and a similar method is applied to GPU for sorting. Reference [5] proposes matrix sort, which is actually an implementation version of parallel merge sort. Reference [7] describes the design of high-performance parallel sorting and merge sorting routines for multi-core GPU. Reference [12] uses OpenCL to accelerate odd-even merge sort on GPU. Reference [13] optimizes the worst-case input of pairwise merge sorting on GPU.

References ([0][10]) belong to Type III. Reference [0] studies the parallel merging and sorting algorithm on CPU, GPU and other hardware, and achieves good results. Reference [10] studies the design of resource independent parallel algorithm, including parallel merge sort algorithm.

References ([19]) belong to Type IV. Reference [19] provided an efficient parallel algorithm for In-place radix sort on CPU.

References ([15]) belong to Type V. Reference [15] provided a radix sorting parallel algorithm suitable for graphic processing unit (GPU) computing .

References ([16][17][20]) belong to Type VII. Reference [16] compares the performance of open source parallel sorting algorithm based on OpenMP. Reference [17] uses OpenMP to design a parallel bubble sorting algorithm. Reference [20] provided a framework for the automatic vectorization of parallel sort on x86-Based processors.

References ([4][6][9][18]) belong to Type VIII. Reference [4] uses merge path to calculate the visible area in the collection of disjoint segments within the plane.  In Reference [6] a parallel sample sorting algorithm suitable for GPU is proposed. Reference [9] provided a fast parallel GPU sorting method using a hybrid algorithm. Reference [18] it implements the parallel sorting algorithm of data on CUDA platform.

References ([8]) belong to Type IX. Reference [8]  focus on oblivious algorithms for multicores and network of processors.

## 3 CLASSIFICATION OF RESEARCH METHODS

Table 2. Different Research Methods

| Whether to use mixed sorting method | Improvement Strategy | | | | | |
|---|---|---|---|---|---|---|
| | Load Balance | Cache Efficient | Load Balance&Cache Efficient | Reduce/Avoid Bank Conflict | Computation Optimization | Others |
| No | Ⅰ. [4][14] | Ⅱ. [8] | Ⅲ. [0][1][2][19] | Ⅳ. [6][13][21] | Ⅴ. [3][5][15] | Ⅵ. [12][17] |
| Yes | Ⅶ. [7][9] | Ⅷ. [10] | Ⅸ. | Ⅹ. | Ⅺ. [11][20] | Ⅻ. [16][18] |

### 3.1 Criteria

Most classical sorting algorithms are designed on the premise of serial execution, while the sorting network represented by dual tone sorting and parity sorting is a sorting method specially designed for parallel machines. The sorting network is composed of multiple comparators connected according to certain logic, and the position and comparison results of each comparator are clearly determined, There will be no branch prediction of the program, so it is very suitable to use modern parallelization technologies, such as single instruction multi data (SIMD) flow technology.

At the same time, although the sorting network such as Bi tonal sorting was originally designed for parallel machines, it has o (n × Log2n). Therefore, researchers have proposed many parallel sorting results to reduce the time complexity. In theory, the time complexity can reach o (n × Log n), but due to the high communication cost in the sorting process and the difference between the theoretical model and the actual model, the actual time complexity should reach o (n × Log n) is very difficult. Only part of the objectives of these research results can be achieved in the actual parallel machine environment. The performance acceleration effect is not obvious than the serial sorting algorithm, so corresponding improvement and optimization work needs to be done.

Under the condition of modern CPU and new hardware processor, the optimization of sorting operation mainly starts from the optimization implementation of sorting method. Based on the hardware patterns of the equipment, considering how to combine the patterns of the method itself with the structural patterns of the hardware, and using the parallel patterns of multi-core and multi-threaded processing of the equipment, the parallelism of all levels of the algorithm is improved; At the same time, considering the storage access level of the device, the high-speed storage unit on the device is used to reduce the memory access delay, optimize the arrangement of data in the storage, and improve the memory access efficiency of the method.

In this section, two independent and different criteria would be used to divide research objects into different types:

1) **Whether to use mixed sorting method**. There are two types here: **No** or **Yes**.

2) **Improvement Strategy**. There are six kinds of improvement strategy here: **Load Balance, Cache Efficient, Load Balance&Cache Efficient, Reduce/Avoid Bank Conflict, Computation Optimization, Others.** Load balancing is to distribute computing tasks evenly to each thread (core) as far as possible, so as to minimize the barrel effect. Cache efficient is to make efficient use of the cache, especially in the face of a large amount of data, try to put the commonly used data at a certain time in the memory with fast access speed. Reduce/Avoid Bank Conflict is to reduce or avoid the probability of memory conflict as much as possible, because in some algorithms, the cost of memory conflict is high, which will bring a series of effects, so as to reduce the efficiency of sorting algorithm. Computation Optimization is a way to improve the performance of the whole algorithm by optimizing the calculation process of the algorithm, such as reducing the computational complexity of the algorithm as much as possible.

### 3.2 The Classification

Based on the appeal classification standard, we give the classification in Table 2. The meaning of each class is as follows:

**Type I.** This type uses a single sorting method and adopts load balancing strategy in parallel algorithm for optimization.

**Type II.** This type uses a single sorting method and uses efficient cache strategy for optimization in parallel algorithms.

**Type III.** This type is optimized by using a single sorting method, load balancing and efficient cache strategy in parallel algorithms.

**Type IV.** This type uses a single sorting method and adopts the strategy of reducing / avoiding memory conflict in parallel algorithm for optimization.

**Type V.** This type uses a single sorting method and adopts the calculation process optimization strategy in the parallel algorithm.

**Type VI.** This type uses a single sorting method, and uses other optimization strategies or no obvious optimization strategies in parallel algorithms.

**Type VII.** This type uses hybrid sorting method, and uses load balancing strategy in parallel algorithm to optimize parallel algorithm.

**Type VIII.** This type uses hybrid sorting method, and uses efficient cache strategy in parallel algorithm to optimize.

**Type IX.** This type uses hybrid sorting method, and adopts load balancing and efficient cache strategy in parallel algorithm to optimize.

**Type X.** This type is optimized by using hybrid sorting method and reducing / avoiding memory conflict in parallel algorithm.

**Type XI.** This type uses hybrid sorting methods and uses computational process optimization in parallel algorithms.

**Type XII.** This type uses mixed sorting methods, and uses other optimization strategies in parallel algorithms, or there is no obvious optimization strategy.

Explanation of Different Types

References ([4][14]) belong to Type I. Reference [4] uses cascading divide and conquer technology and mergepath to average the workload between processors. Reference [14] proposes a parallel merge sort algorithm with load balancing strategy.

References ([8]) belong to Type II. In reference [8], Aiming at several basic problems such as matrix transpose, FFT, sorting, Gaussian elimination normal form, list sorting and connected components, a mo (multicore oblivious) algorithm using cache efficiently is proposed.

References ([0][1][2][19]) belong to Type III. Reference [0][1][2] proposes a new parallel segmentation method and an original data structure: merge path and merge matrix It should be noted that the segmentation results and computational complexity are the same as those of previous algorithms, but the segmentation methods are different and enlightening. Based on the above, a synchronization free, cache efficient merging algorithm is proposed. Reference [19] present a novel parallel in–place radix sort algorithm, PARADIS, which addresses both problems: a) "speculative permutation" b) "distribution–adaptive load balancing".

References ([6][13][21]) belong to Type IV. Reference [6] makes efficient use of the multi–level memory structure of GPU, and points out that the general multiplexing technology and especially sample sorting achieve better performance than bidirectional merge sorting and fast sorting. In reference [13], A series of strategies are proposed to optimize the memory access serialization problem caused by bank conflict in parallel sorting on GPU, such as memory alignment, general strategy and so on. Reference [21] proposes an optimal parallel merge and sort without memory conflict.

References ([3][5][15]) belong to Type V. Reference [3] proposes a parallel merging and sorting method based on complete binary tree to reduce the computational complexity of the algorithm. Reference [5]

An efficient parallel algorithm for merging and sorting is proposed: GPU matrix sort. Matrix sort mainly has two steps: 1. Sort the rows of the matrix; 2. Recursively merge the two sorted rows. Reference [15] used advantage of the parallelism of GPU numerical computing processing, a parallelization design method of quadratic least significant number (LSD) first cardinal sorting (b_lsd_rs) algorithm based on open computing language (OpenCL) is proposed

References ([12][17]) belong to Type VI. Reference [12] proposes a novel implementation of parallel odd–even merge sort based on the recent GPU OpenCL programming model, based on Knuth's algorithm, and makes some changes. Reference [17] proposes an implementation method of parallel bubble sorting.

References ([7][9]) belong to Type VII. Reference [7] this paper describes the design of high-performance parallel sorting and merge sorting routines for multi-core GPU, in which load balancing strategy is widely used. Reference [9] proposes a hybrid scheduling method based on load balancing strategy.

References ([10]) belong to Type VIII. Reference [10] interweaves merge sort and sample sort, and introduces some techniques to reduce the worst-case overhead caused by error sharing.

References ([11][20]) belong to Type XI. Reference [11] reduces the computational complexity of the whole algorithm by performing piecewise parallel quick sorting on the data, and then performing recursive merge sorting. Reference [20] proposes a multi-core parallel automatic vector sorting framework to optimize the computing process from the bottom.

References ([16][18]) belong to Type XII. Reference [16] mainly studies the linear parallelism of sorting algorithm, and does not involve the optimization strategy. Reference [18] uses CUDA to design parallel sorting algorithm on GPU, and does not involve optimization strategy.

## 4  REVIEW OF EXPERIMENTAL ANALYSIS

In this section, we will classify the metric of evaluation and system parameters, as shown in Table 3. In Table 3, all experimental analysis is also classified according to the metric and parameters. It can be seen from Table 3 that most of the references compare speedup, time, sorting rate, bank conflict per element and others.

Table 3. Experiments with Different Metric and Parameters

| Parameters | Metric | | | |
|---|---|---|---|---|
| | Speedup | Time | Sorting Rate | Bank Conflicts per element |
| Number of threads/blocks | I. [0][1][2][11][14][16] | II. [2][16][18][19] | III. | IV. |
| Data | V. [0][2][4][11][12][14][15][17][20] | VI. [2][5][7][9][14][15][17][18][19][20] | VII. [6][7][13] | VIII. [13] |
| Method/Implementation | IX. [2][11][15][16][20] | X. [2][7][9][15][16][19][20] | XI. [6][7][13] | XII. |
| System | XIII. [12][14] | XIV. [7][14] | XV. [7] | XVI. |

### 4.1 Metric of Evaluation

**Speedup** means that the speed improvement ratio of parallel algorithm is higher than that of single core serial algorithm. The formula is as follows:

$$\text{Speedup} = \frac{\text{Time}_s}{\text{Time}_p}$$

**Sorting** rate means number of elements sorted per second. The formula is as follows:

$$\text{Sorting Rate} = \frac{\text{Data Size}}{\text{Time}}$$

**Bank conflict per element** means the average number of memory conflicts per element. The formula is as follows:

$$\text{Bank Conflict per Element} = \frac{\text{Number of Bank Conflict}}{\text{Data Size}}$$

Other metric includes **Time**, **Others**.

### 4.2 System Parameters

**Number of threads/blocks** represents number of computing units participating in parallel sorting.
**Data Size** represents the size of the data quantity.
**Data Type** represent the type of data, such as integer, floating point, vector, etc.

**Data Pattern** represent the pattern of the data, such as the random, even/odd, pipe organ, sorted, push front, etc.

**Method/Implementation** represent different sorting methods, or different versions of implementations.

**System** represent the system platform used to complete the experiment.

### 4.3 Experimental Comparison

·**Speedup**

Speedup is an indicator that many researchers pay attention to. It can intuitively measure the improvement and promotion of parallel algorithm compared with traditional serial algorithm. The researchers compared the speedup of the algorithm under different parameters, such as **number of threads/blocks** in reference [1][2][11][14][16]. By comparing the speedup ratio of algorithms with different thread numbers, we can intuitively reflect the parallelism and performance of parallel algorithms. Another common parameter is the **data size**. By comparing the algorithm acceleration ratio under different amounts of data, we can see that the parallel algorithm performs better than the serial algorithm. There are reference [2][4][11][12][14][15][17][20]. This is a very basic and important indicator and parameter pair. Another parameter is **data type**, which is a relatively specific parameter, and few researchers pay attention to it, where in reference [2][4]. Reference [2][11][15][16][20] also compares the acceleration ratio of algorithms under different **methods / implementations**, which can intuitively reflect the performance difference between their own algorithms and others. In addition, some researchers want to know the performance of the algorithm on different systems, so they choose the system as the parameter. References like this are reference [12][14].

·**Time**

The running time of the algorithm is the simplest measure the performance of the algorithm, which is considered in most of the work, and its parameters are the most extensive. Reference [2][16][18][19] compare the running time of algorithms under different running threads / blocks, so as to test the degree of parallelism. If time is used as a measure, the amount of data is an essential consideration, because for a specific algorithm, its running time is inseparable from the amount of data to be sorted. Almost all time measurement work takes this into account, we can find it in the references [2][5][7][9][14][15][17][18][19]. Another parameter is data pattern, there is only one reference for this parameter, in reference [20]. Similar to the algorithm speedup ratio, many works also pay attention to comparing the algorithm running time under different methods / implementations to measure the advantages and disadvantages of their own work, which is also a common and basic measurement method. We can find it in reference [2][7][9][15][16][19][20].

·**Sorting Rate**

The definition of sorting rate can be found in the previous article. This is a measure of the absolute speed of sorting (the number of sorting elements per unit time) by comprehensively considering the amount of data and sorting time. By comparing this metric, we can more intuitively see the sorting performance of different algorithms and the same algorithm under different parameters. Such consideration is given to reference [6][7][13]. Among them, reference [6] and [7] both compared number of threads/blocks, data type, method/implementation. Reference [13] compares sorting rates under different data sizes and different implementation versions as well as different data pattern.

·Bank Conflict per Element

Bank conflict is a memory conflict, it means that in a shared memory system, multiple processing units access the same information block in the memory at the same time. This conflict is more common with GPUs than with CPUs. In the literature I selected, only one systematically explored this index, that is reference [13]. It compares bank conflicts under different data sizes and different data patterns.

·Others

Most of the literatures in this part do not have obvious measurement methods, or adopt more specific measurement methods not listed above. Due to the limited space, it will not be repeated.

## 5 DISCUSSION AND SUGGESTION

This paper discusses the research methods and research objects of various references and finds that although in the past few decades, researchers have been paying attention to the sorting field, put forward many sorting methods, and made corresponding improvements according to the development of computer hardware and software to improve the performance of sorting methods, there are still many challenging problems to be further studied. Therefore, this paper puts forward the following directions, which can provide directions for future parallel sort algorithm research:

### 1) Load balancing strategy (data partition strategy)

For multi-threaded parallel sorting algorithms, data partition processing is essential. The existing data partition strategies are roughly divided into three types: the first is to distribute data evenly according to the number of threads, the second is to divide data buckets according to certain strategies, and the third is to calculate the distribution of data. Divide according to the distribution. The first division method is relatively simple to implement, with high division efficiency and load balance among threads. Each thread sorts the local data and finally merges the data of each thread. However, this division method does not make use of the patterns of data, resulting in a cumbersome final merging process; The second division method is commonly used in sorting strategies such as cardinality sorting, which require bucket division. After data bucket division, there is an orderly relationship between buckets. In the final merging stage, only the data of each bucket needs to be connected, but the bucket division process is not controllable, resulting in different sizes of each bucket and possible load imbalance; The third partition strategy can first obtain the data distribution, so as to determine the amount of data to be processed by each thread and obtain the load balance of threads, but it also increases the cost of calculating the data distribution. The worst result of calculating the data distribution is that it is still evenly distributed according to the number of threads, wasting the calculation time. The three partitioning strategies have their advantages and disadvantages. How to avoid the worst case of various partitioning strategies to the greatest extent, highlight their advantages, and improve the overall sorting performance will be a problem worthy of study

**2) Optimizing the merging process.**

In the process of sorting, in order to make use of the high-speed access of cache, the input sequence will be divided according to the lowest cache size. When the lowest cache is small or the input sequence is large, a large number of subsequences will be generated, The merging process will inevitably involve a large number of thread communication and data rearrangement. If the data is divided according to a certain size order, the merging process will be simplified. If this idea can be interspersed in the sorting process, In addition, improving the utilization of threads in the merging process, especially when the number of sequences is less than the number of threads, reducing the idle rate of threads without affecting the efficiency of a single thread is also a way to improve the merging efficiency.

**3) Study the pattern of input sequences.**

When the input sequence has some patterns, such as conforming to a specific distribution, skewed data, most of the sequence is orderly, using the patterns of these input sequences is very helpful to simplify the sorting steps. The simpler detection method is to scan the sequence, but when the sequence is large, the cost of detection can not be ignored, Moreover, it is not guaranteed that each detection can get a sequence with certain patterns, which will waste the detection time. How to balance the detection cost and detection efficiency, so that the use of input sequence patterns can really simplify the steps of the sorting process, so as to improve the performance of the sorting algorithm, which is a direction worthy of research.

**4) Sorting of multiple comparison criteria.**

The sorting process will compare the data according to the computer's own data size standard. When the comparison standard becomes complex or different from the computer's own comparison standard, it is necessary to convert the comparison standard into a standard that can be recognized by the computer. When the required comparison standard is complex or various comparison standards need to be considered, It will be a time-consuming process to convert each pair of comparison data into comparison standard. How to optimize this conversion process and minimize the impact on sorting performance is an interesting problem.

# 6 CONCLUSIONS

Sorting is a very important problem in the computer field. Sorting technology will be used in many operations. This paper studies the parallel memory sorting method on modern hardware, and summarizes its research status and progress through the research problems, research methods and measurement methods of the target literature and references. In these three dimensions, the work content is classified and made into three tables. The paper also gives the innovative direction of future research such as data partition strategy, optimizing the merging process, study the pattern of input sequences and sorting of multiple comparison criteria.